\documentclass[twocolumn]{IEEEtrans}
\usepackage{graphicx}
\usepackage{latexsym}

\def\BibTeX{{\rm B\kern-.05em{\sc i\kern-.025em b}\kern-.08em
    T\kern-.1667em\lower.7ex\hbox{E}\kern-.125emX}}

\begin{document}

\title{Simulated Annealing Approach to the Temperature-Emissivity Separation Problem in Thermal Remote Sensing Part One: Mathematical Background}

\author{John A. Morgan\\The Aerospace Corporation\\P. O. Box 92957\\Los Angeles, CA 90009 \\ john.a.morgan@aero.org}

\maketitle

 \begin{abstract}The method of simulated annealing is adapted to the temperature-emissivity separation (TES) problem. 
 A patch of surface at the bottom of the atmosphere
is assumed to be a greybody emitter with spectral emissivity $\epsilon(k)$ describable by a mixture of spectral endmembers. We prove that a simulated annealing search conducted according to a suitable schedule converges to a solution maximizing the \emph{A-Posteriori} probability that spectral radiance detected at the top of the atmosphere originates from a patch with stipulated $T$ and  $\epsilon(k)$. Any such solution will be nonunique. The average of a large number of simulated annealing solutions, however, converges almost surely to a unique Maximum A-Posteriori solution for $T$ and $\epsilon(k)$. 

The limitation to a stipulated set of endmember emissivities may be relaxed by allowing the number of endmembers to grow without bound, and to be generic continuous functions of wavenumber with bounded first derivatives with respect to wavenumber.

\end{abstract}


\section{INTRODUCTION}

\PARstart{T}he temperature-emissivity separation (TES) problem bedevils any attempt to extract spectral information from remote sensing in the thermal infrared. 
A variety of methods has been proposed for handling the temperature-emissivity separation (TES) 
problem~\cite{Dash2002,LiEtAl2013}.  In most of them, simultaneous LST and band emissivity retrieval 
depends upon specifying an emissivity value in one or more reference bands.  The MODIS Land Surface Temperature
(LST) algorithm~\cite{Wan1999} seeks a pair of reference channels in a part of the 
thermal spectrum in which the emissivity of natural surfaces displays very limited variation, and may therefore be regarded 
as known with good confidence.  Multiband emissvities inferred on this basis are called "relative" 
emissivities~\cite{Li1999}.  Other algorithms of this nature include the reference channel 
method~\cite{KahleAlley1992}, emissivity normalization~\cite{KealyHook1993},  temperature-independent spectral index 
method~\cite{PetitcolinVermote2002},~\cite{LiBecker1993} and spectral ratios~\cite{Watson1992}.  
The study by Li et al.~\cite{Li1999} shows that all of these relative emissivity retrieval algorithms are closely related, and argues that 
they may be expected to show comparable performance.  The
analysis of Multispectral Thermal Imager (MTI) data~\cite{BorelSzymanski1998} depends on collection of radiance from a surface with 
looks at nadir and 60 degrees off-nadir, assuming a known angular dependence of emissivity, in order to balance equations and unknowns.  
The generalized split-window LST algorithm~\cite{WanDozier1996} likewise uses dual looks in a regression-law based approach. 
The "grey body emissivity" approach~\cite{BarducciPippi1996} exploits the slow variation of emissivity with wavelength for certain 
natural targets, while the physics-based MODIS LST algorithm~\cite{WanLi1997} exploits observations taken at day and at night, on the assumption that band 
emissivites do not change over periods of a few weeks.  A study with the Airborne Hyperspectral Scanner \cite{SobrinoEtAl2006} compares multiple
TES approaches.
 
We shall investigate a simulated annealing approach to the TES problem. The approach is an extension of earlier work on Bayesian TES \cite{Morgan2005,Morgan2012}. Simulated annealing cannot give a unique solution to this problem, but we shall prove that the average of a large number of simulated annealing TES solutions converges almost surely to a unique TES estimate.

This paper will concentrate on the mathematical basis of the algorithm and a proof of its convergence. A study of its performance will form the subject of a subsequent paper.

\section{BACKGROUND}

Simulated annealing has traditionally been regarded as a preferred method of global solution for combinatorial optimization problems such as Traveling Salesman. In this paper, we adapt the Metropolis algorithm \cite{MetropolisEtAl1953,KirkpatrickEtAl1983,Press1988} to an optimization problem that lacks a unique global optimal solution: Temperature-emissivity separation. The underdetermined temperature-emissivity separation (TES) problem, notoriously \cite{Dash2002}-\cite{SobrinoEtAl2006}, has a continuous infinity of solutions that yield the identical optimum value for any cost or payoff function one cares to choose.

A key part of any simulated annealing algorithm is the choice of an annealing schedule that causes the posterior probabilities to transition from nearly uniform to very tight in such a way as to evade the risk of the MAP search from converging to a local, rather than a global, optimum. The look and feel of the justification for this approach is ergodic.

In what follows we shall mostly concern ourself with the existence of a solution to the simulated annealing TES problem, and shall simply assume that a suitable annealing schedule has been supplied. Factors that enter into the choice of annealing schedule are described in \cite{KirkpatrickEtAl1983,Press1988}. Selection of the annealing schedule and sample TES retrievals will be the subject of a subsequent paper.
\section{Simulated Annealing and the temperature-emissivity separation problem}

 \subsection{Metropolis Algorithm Search for Maximum A-Posteriori Solution}

Suppose that we have in our possession prior knowledge that a target patch that forms part of the lower boundary of the atmosphere is composed of an intimate mixture of 
$m+1$ spectral endmembers $\{\epsilon_{i}(k)\}$ at temperature $T$. 
For later convenience, we shall require that spectral emissivities be bounded continuous functions of wavenumber with
bounded first derivative with respect to wavenumber. The label $k$ may, depending upon context, refer to wavenumber, or to a finite number of wavenumber-averaged
spectral bands. Except in Section \ref{sect:arbitrary}, we shall assume the band interpretation. 

The spectral mixture amounts to a mapping into a geometric m-simplex whose vertices have spectral endmembers at a stipulated temperature $T$ for coefficients. Suppose we have 
$m+1$ distinct points $\mathbf{y}_{0}, \mathbf{y}_{1}, \cdots \mathbf{y}_{m}$ in $\mathbf{{R}^{m}}$ chosen so that the vectors 
$\mathbf{y}_{1}-\mathbf{y}_{0}, \mathbf{y}_{2}-\mathbf{y}_{0}\cdots \mathbf{y}_{m}-\mathbf{y}_{0}$ are linearly independent. Then the set
\begin{equation}
 K_{m} \equiv \sum_{i=0}^{m} \lambda_{i} \mathbf{y}_{i}
 \end{equation}
 with
\begin{equation} 
\lambda_{i} \ge 0, \forall \, i
\end{equation}
and
\begin{equation} 
\sum_{i=0}^{m} \lambda_{i}=1
\end{equation}
is an m-simplex.~\cite{Rotman1988}
A spectral mixture with stipulated weights $\lambda_{i}$ corresponds to the vector~\footnote{Should the target patch contain an isothermal checkerboard mixture of end members, the weight $\lambda_{i}$ is to be interpreted as the fraction of the total
surface area of the patch occupied by the $i$-th subregion
$\lambda_{i}=\frac{A_{i}}{\sum_{j=1,n} A_{j}}$.
The product $\lambda_{i} \epsilon_{i}(k)$ is thus a normalized emissivity-area product for that subregion. We defer the case of endmembers with differing
temperatures to a later date. We believe it, however, to be a straightforward extension of the reasoning in this paper.}
\begin{equation}
\mathbf{x}=\sum_{i=0}^{m} \lambda_{i} \mathbf{y}_{i} \in \mathbf{R}^{m}.
\end{equation}


The \emph{interior} of $K_{m}$ is the subset of $K_{m}$ for which $\lambda_{i} > 0$, that is the closure of its interior. 
The polyhedron of $K_{m}$, denoted $|K_{m}|$, is the set comprised of the points of  $\mathbf{x} \in K_{m}$ considered as a subset of  $\mathbf{{R}}^{m}$, and is a convex compact 
subset of $\mathbf{{R}^{m}}$.

In the case $m+1=3$, a familiar example of a 2-simplex is the ternary diagram used to classify phreatic igneous rocks. 
The double three-component diagram used in the QAPF classification~\cite{LeBasStreckeisen1991} scheme is a union of two 2-simplices, and is an example of a simplicial complex. 

For present purposes, the $i^{th}$ pure endmember for the $n^{th}$ trial is assigned to the $i^{th}$ vertex of $K_{m}$
\begin{equation}
\mathbf{y}_{i} \Leftrightarrow \epsilon_{i }(k) , 0 \le i \le m
\end{equation}
with the spectral mixture corresponding to a point in the polyhedron of $K_{m}$,

It is necessary to account for surface temperature in a somewhat different way. Let the minimum and maximum physically admissible surface temperatures be $T_{min}$ and
$T_{max}$, respectively. Then the temperature of our target patch is given by
\begin{equation} 
T=(1-\lambda_{m+1}) T_{min} + \lambda_{m+1} T_{max} \label{eq:T_homotopy}
\end{equation}
with
\begin{equation} 
0 \le \lambda_{m+1} \le 1.
\end{equation}
Corresponding to $\mathbf{x}$ introduced already, we have from Eq. (\ref{eq:T_homotopy}) 
\begin{equation}
x_{m+1} \in I^{1}, 
\end{equation}
the unit interval, with 
\begin{equation}
x_{m+1} \Leftrightarrow T_{n}
\end{equation}

The quantity that appears in the forward model for the $n^{th}$ trial is
\begin{equation}
\left < \epsilon(k) B_{k}(T_{n}) \right > =\sum_{i=1}^{m} \lambda_{i}\epsilon_{i }(k) B_{k}(T_{n}). \label{eq:mixed} 
\end{equation}
$B_{k}(T_{n})$ is the (band-integrated, as needed) Planck function at temperature $T_{n}$. The parametrization of the choice $\{T_{n},\epsilon(k)\}$ 
in terms of the vector $x$ is a mapping into the topological product
\begin{equation}
H^{m+1} \equiv I^{1} \otimes \left |K_{m} \right |
\end{equation}
of $I^{1}$ and $\left |K_{m} \right |$.
The set $H^{m+1}$ is not a simplex, nor is it necessarily a simplicial complex. It is, however, a convex polytope, and is the convex hull of its vertices
$\mathbf{x}_{i}, 0 \le i \le m+1$.~\footnote{Although we will not need it in what follows, $H^{m+1}$ can be decomposed into either a simplicial complex or a union
of simplices.}


We score trial mixtures by that we most wish to maximize: The posterior
probability for the observed spectral radiance to originate from a surface patch with temperature $T$ and spectral emissivity $\epsilon(k)$.
A standard argument~\cite{Morgan2005,Morgan2012} gives the posterior probability in terms of a MAXENT estimator
\begin{equation}
P(I\mid T,\epsilon,\sigma)=
exp\left[-\frac{(I-I_{FM})^{2}}{2\sigma^{2}(T_{a})}\right]\frac{dI}{\sigma(T_{a})} \label{eqn6}
\end{equation}
in terms of a 
forward model
\begin{equation}
I_{FM}=f \left (\sum_{i=1}^{m} \lambda_{i}\epsilon_{i }(k) B_{k}(T_{n}) \right) \Leftrightarrow f(x) 
\end{equation}
that is some function of the $n^{th}$ trial,
in each spectral bin $k$. We note that while the equation of transfer is linear, the dependence of its solution $I_{FM}$ upon $\epsilon_{i }(k) B_{k}(T_{n})$ need not be.
The assumed noise variance $\sigma^{2}$ is shown as having a formal dependence upon a parameter, the "annealing temperature" $T_{a}$, which 
governs the annealing schedule for the search for a Maximum A-Posteriori solution. The joint posterior probability in $J$ spectral bands is proportional to
\begin{equation}
P(\{I_{k}\} \mid T,\epsilon,\sigma)=
\prod_{k=1}^{J} \, exp\left[-\frac{(I_{k}-I_{FM}(k))^{2}}{2\sigma^{2}(T_{a})}\right]\frac{dI}{\sigma(T_{a})} \label{eqn7}
\end{equation}

If radiance $I_{k}$ in each of J bands originating from a patch on the Earth's surface has been detected at the top of the atmosphere (TOA),  
the posterior probability that the surface patch is at a temperature T given prior knowledge $K$ is given by Bayes' theorem as 
\begin{equation}
P(T, \epsilon_{i}(k)  \mid \{I_{k}\},K)=
P(T,  \epsilon(k) \mid K)\frac{P(\{I_{k}\} \mid T, \epsilon_{i}(k), K)}{P(\{I_{k}\}\mid K)}. \label{eqn3}
\end{equation}
The noise variance is assumed known and the functional dependence of probabilities upon $\sigma_{i}$ is omitted.
The prior probability $P(\{I_{i}\}\mid K)$ for the radiances $\{I_{k}\}$ has no dependence upon $T$ and for our purposes may be absorbed into an overall normalization.~\cite{Bretthorst1988} Equation (\ref{eqn3}) is evaluated with aid of 
the prior probability for the surface to be at temperature T and have spectral emissivity $\epsilon(k)$, given available knowledge $K$ \cite{Morgan2005},
\begin{equation}
P(T,  \epsilon(k) \mid K) \, dT \,  \propto \prod_{k} d \epsilon(k) \frac{dT}{T}.  \label{eq:prior}
\end{equation}
$P(T, \epsilon_{i}(k) \mid \{I_{i}\},K)$ is the conditional probability for the hypothesis that the surface temperature is $T$, and the spectral emissivity $\epsilon_{k}$, given
observed radiances $\{I_{i}\}$ and prior knowledge $K$.

Each trial is thus scored according to the joint posterior probability for observed spectral radiance $I_{i}$ to result from surface temperature $T$ and spectral emissivity $\epsilon_{k}$,
\begin{equation}
p_{n} =P(T_{n}, \epsilon(k) \mid \{I_{i}\},K) \equiv p_{n}(x)  \label{eq:posterior}
\end{equation}
where $x$ stands for $\{\mathbf{x},x_{m+1}\}$.
Thus, in going from the $(n-1)^{th}$ to the  $n^{th}$ trial, the $n^{th}$ candidate mixture is selected by Monte Carlo draw and $p_{n}$ for the new trial is compared to 
$p_{n-1}$ for the last one. The probability that it is accepted is~\cite{MetropolisEtAl1953,Press1988}
\begin{equation}
P = \left \{
 \begin{array}{ll}
1 & \mbox{if $p_{n}/p_{n-1} \ge 1$} \\ \label{eq:Metropolis}
P(T_{a}) & \mbox{otherwise}
\end{array}
\right . 
\end{equation}
where the probability $P(T_{a})$ of taking a downward step in $p_{n}$ is determined by the annealing schedule. The dependence of $P(T_{a})$ on the annealing schedule 
is symbolized by the annealing "temperature" $T_{a}$ which is taken to decrease systematically during the MAP search. The actual form $P(T_{a})$ takes in practical calculations is determined empirically.

\subsection{Convergence} \label{sect:convergence}

\subsubsection{Spectral mixtures comprising a finite number of endmembers} \label{sect:finitecase}
We now examine the question of convergence.
Corresponding to the sequence of m-simplices $K_{m}$ as the number of trials $n$ 
increases without bound is a sequence of trials $\{T_{n},\epsilon_{i}(k)\}$ with associated loci $\{x\} \in H^{m+1}$. 

As a closed bounded subset of $\mathbf{R}^{m+1}$, $H^{m+1}$ is a compactum. Therefore, as $n \rightarrow \infty$, the sequence of trials  $x$ contains a convergent subsequence, whatever the value of $m$. Correspondingly, the sequence of posterior probabilities likewise has a convergent subsequence that, by construction, tends to the maximum value of the posterior probability, i.e., to a MAP solution for $T$ and $\epsilon(k)$.

Consider the map $x'=\Phi(x)$ given by
\begin{equation}
\Phi(x) = \left \{ \begin{array}{ll}
x' & \mbox{if $p_{n}(x')-p_{n}(x) \ge 0$} \\ \label{eq:Nextmap}
x & \mbox{otherwise.}
\end{array}
\right .
\end{equation} 
The mapping Eq. (\ref{eq:Nextmap}) gives the action of the Metropolis algorithm according to Eq. (\ref{eq:Metropolis}) at sufficiently late times in the
annealing schedule that a transition to a state of decreased posterior probability occurs rarely; in the limit, almost never.
We have noted that at a sufficiently late point in the annealing 
schedule, trials that decrease the posterior probability Eq. (\ref{eq:posterior}) will become infrequent. We may elide any such trials without affecting the 
convergence of the subsequence, which then takes the form
\begin{equation}
x_{n+1}=\Phi(x_{n})
\end{equation}
For all $n$ greater than some $M$, convergence of the subsequence implies the Cauchy condition 
\begin{equation}
d(x_{n},x_{n+1})= d(x_{n},\Phi(x_{n}))< \epsilon,
\end{equation}
(with the Euclidean norm supplying a suitable metric for finite $m$) so that 
\begin{equation}
x \rightarrow \Phi(x). \label{eq:PhiFP}
\end{equation}
The mapping Eq. (\ref{eq:Nextmap}) generates a sequence of trials $x$ for which $p_{n}$ is nondecreasing. By Zorn's Lemma, the set comprised of all admissible trials  $x$ has at least one element with a maximal value of  $p_{n}$. We note that maximizing $p_{n}$ also maximizes the information-theoretic entropy
by Eq. (\ref{eqn6}). According to the usual statement of the Second Law, the state of maximum entropy is one of thermodynamic equillibrium. We may therefore, in a nod to Refs. \cite{MetropolisEtAl1953} and (\cite{Nash1951} both, call the limit Eq. (\ref{eq:PhiFP}) an equillibrium point. 

We note that, in the limit, Eq. (\ref{eq:PhiFP}) amounts to a fixed point of Eq. (\ref{eq:Nextmap}). Ordinary fixed-point theorems are inapplicable to Eq. (\ref{eq:Nextmap}) because it is neither continuous nor semicontinuous: It can map an open set $\in H^{m+1}$ to a singleton $x'$. We can, however, adapt a celebrated construction introduced by Nash~\cite{Nash1951} to prove the existence of a fixed point of an equivalent self-mapping. 

In fact, we shall prove a somewhat stronger result.
Consider
\begin{equation}
\phi_{\alpha}=max(0,p_{n}(x_\alpha)-p_{n}(x)).
\end{equation}
for stipulated $x$. The function $\phi$ is continuous in the mixture $x_\alpha$.  Define the mapping 
$N:x \rightarrow x'$ by
\begin{equation}
x'=\frac{x+\sum_{\alpha} \phi_{\alpha} x_{\alpha}} 
{1+\sum_{\alpha} \phi_{\alpha}},  \label{eq:Nashmap}
\end{equation}
where the index $\alpha$ is taken to run over members of any finite set of admissible trials $x_\alpha$
in the execution of the Metropolis algorithm. (One may think of the collection of all sequences $x_\alpha$ in ensemble-theoretic
terms.) 
Suppose that $x'$ is a fixed point under  Eq. (\ref{eq:Nashmap}).
In Eq. (\ref{eq:Nashmap})
some values of $\alpha$ correspond to choices for $\{ T_{i}, \epsilon_{i}(k) \}$ for 
which the posterior probability does not increase:
\begin{equation}
p_{n}(x_{\alpha})-p_{n}(x) \le 0.
\end{equation}
For these values of $\alpha$, 
\begin{equation}
\phi_{\alpha}=0.
\end{equation}
If the choice $x$ is fixed under the mapping $N$ in Eq. (\ref{eq:Nashmap}), then the
contribution to $x'$ from any $x_{\beta}$ must not decrease; therefore,
$\phi_{\beta}=0, \forall{\beta}$, lest the denominator in $\Phi$ exceed unity.  Put another way, no
other choice of $\{ T_{i}, \epsilon_{i}(k) \}$ can increase the posterior probability.  
But that is the definition of an equillibrium point.  

If, on the other hand, an equillibrium point $x$ maximizes the posterior probability Eq. (\ref{eq:posterior}), every $\phi_{\alpha}$ vanishes, so that
$x$ is a fixed point. 

Equation (\ref{eq:Nashmap}) is continuous and maps points $x$ into a convex compactum $\subset \mathbf{R}^{m+1}$.    
A fixed point
\begin{equation}
x=N(x) \label{eq:NashFP}
\end{equation}
exists according to the Brouwer fixed-point theorem that, by construction, maximizes the \emph{a-posteriori} probability of $x$.

The mapping Eq. (\ref{eq:Nextmap}) generates a sequence of trials $x$ for which $p_{n}$ is nondecreasing and gives the maximal value of $p_{n}$ in the limit, while Eq. (\ref{eq:NashFP}) demonstrates
the existence of a trial $x^{*}$ for which $p_{n}$ cannot be made greater. 
In view of the ensemble-theoretic freedom to choose $x_\alpha$, we may identify the limit in Eq. (\ref{eq:PhiFP}) with the fixed point in Eq. (\ref{eq:NashFP}). 
Therefore, a convergent subsequence of annealing trials exists that tends to an equilibrium point.
Moreover, Eq. (\ref{eq:NashFP}) demonstrates that the annealing search can, in principle, find $x^{*}$ in a finite number of trials.
We conclude that, granted a suitable annealing schedule, there exists at least one convergent sequence of trials that tends  
to MAP surface temperature and spectral emissivity estimates consistent with observed spectral radiances $I_{k}$. 

\subsubsection{Arbitrary spectral emissivities} \label{sect:arbitrary}

The search algorithm just described assumes that the emissivity $\epsilon(k)$ is describable by a mixture of a finite set of spectral end members. While the spectral mixture characterization of $\epsilon(k)$ is of interest in its own right, it may be considered a stronger hypothesis than is strictly desirable. In particular, it seems intuitively reasonable that the simulated annealing approach to TES should work just as well-and admit a simpler algorithmic realization-by using trials with randomly chosen spectral emissivities, rather than by seeking a spectral mixture from a predetermined set of endmembers.

In fact, it is possible to reduce the case of search using arbitrary $\epsilon(k)$ for trials to an extension of the analysis in 
the preceding section by allowing the number of endmembers $m$ to grow without limit for each trial $n$.
Instead of self-mappings into a single polytope with fixed $m$, we consider a sequence of $H^{m_{n}+1}$ as 
$m_{n} \rightarrow \infty$, for each $n$ in the annealing schedule.

The connection between arbitrary $\epsilon(k)$ and a spectral mixture with whose endmembers are allowed to grow 
without limit is easily seen. If we chose $x_{i}, 0 \leq i \leq m_{j}$ from a set of randomly chosen endmembers
$\epsilon_{i}(k)$, it is clear that any random $\epsilon(k)$ can be constructed as a spectral mixture of other random endmembers. By induction: A single endmember $\epsilon_{1}(k)$ trivially reproduces an arbitrary  $\epsilon(k)$ if it is chosen so $\epsilon_{1}(k)=\epsilon(k)$. Suppose that any $\epsilon(k)$ equals a spectral mixture of $j$ suitably chosen
random $\epsilon_{i}(k)$. Then, by the inductive hypothesis for $j=2$, it is possible to replicate any other $\epsilon(k)$ by
a mixture of some spectral emissivity of $j$ endmembers and a $(j+1)^{st}$ random endmember.

We proceed by constructing the polytope for an arbitrary number of spectral endmembers. The polyhedron of the m-simplex $K_{m}$ with unit diameter may be circumscribed by an m-sphere of radius~\cite{BlumenthalWahlin1941}
\begin{equation}
r \le \sqrt{\frac{m}{2(m+1)}}.
\end{equation}
$|K_{m}|$ is thus a subset of the topological product of n replicas of the unit interval $[0,1]$
\begin{equation}
|K_{m}| \subset I^{m} \equiv [0,1] \otimes  [0,1] \cdots  [0,1]
\end{equation}
Every m-simplex is thus contained within the topological product of a countable infinity of replicas of the unit interval $[0,1]$~\cite{Milnor1957}
\begin{equation}
|K_{m}| \subset I^{\infty} \equiv [0,1] \otimes  [0,1] \cdots, 
\end{equation}
as is every convex polytope
\begin{equation}
H^{m+1} \subset I^{\infty}. 
\end{equation}
All the polyhedra $|K_{m}|$ and polytopes $H^{m+1}$ are compact, and by Tychonoff's theorem, the set $I^{\infty}$ which circumscribes every  $|K_{m}|$ and $H^{m+1}$ is
likewise sequentially compact.

$I^{\infty}$ is homeomorphic to the \emph{Hilbert cube}. The Hilbert cube
\begin{equation}
\mathcal{H}  \equiv [0,1] \otimes  [0,\frac{1}{2}] \cdots  [0,\frac{1}{n}]  \cdots
\end{equation}
is a subset of a Hilbert space with the $\textit{l}_{2}$ norm.\footnote{For any finite dimensional subspace of $\mathcal{H}$ however, we may still take the Euclidean norm when choosing a metric.}
$I^{\infty}$is therefore a \emph{complete} space: The sequence of trials, by sequential compactness of the Hilbert cube, 
and sets homeomorphic to it, possesses a convergent subsequence whose limit is, by completeness of $I^{\infty}$, an element of that space. The limit of the convergent subsequence of m-polyhedra is likewise contained within $I^{\infty}$. 

The treatment of convergence in Section \ref{sect:finitecase} requires modification when the number of spectral endmembers is allowed to grow without
limit. At each $n$, the self-mapping $N$ and function $\phi(x)$ are applied to $H^{m_{n}+1}$ as before to give existence of an equilibrium fixed point. 
The simplex dimension $m_{n}$ is allowed to grow without bound, however. The resulting sequence of equilibrium points $\in H^{m_{n}+1}$ possesses a convergent subsequence $\subset H^{m_{n}+1}$ for each value of $n$. 

With that caveat, as $n \rightarrow \infty$ 
 the sequence $\{ x_{n} \}$ converges to a MAP estimate of $T$ and an arbitrary $\epsilon(k)$ by the same reasoning used in the previous section.
In consequence, the limiting equilibrium point of the convergent sequence $\Phi(x_{n})=x_{n}$ as $m_{n} \rightarrow \infty$ will tend to an estimate
of the MAP value for $\{T,\epsilon_{i}(k)\}$ for any admissible $\epsilon(k)$.
We conclude that a suitable simulated annealing search will converge to an arbitrary spectral emissivity that gives 
a MAP estimate of $\{T,\epsilon_{i}(k)\}$.

\subsection{Uniqueness} \label{Uniqueness}

Whatever the dimensionality of the spectral endmember parameterization of emissivity, sequential compactness guarantees existence of a convergent 
subsequence of trials. In practice, we must expect that there will be more than one such sequence.
The nonuniqueness of solutions to the TES problem suggests that there will be a continuous infinity of possible trials $\{T_{n},\epsilon_{i}(k)\}$ that yield any
stipulated value for the posterior probability. In any realizable search strategy, however, we need only contend with a countable set of convergent subsequences. Amongst these 
there will be one for which the posterior probability is greatest.\footnote{One may appeal to Zorn's lemma again at this point, if desired.} This will be the closest approach
to the Maximum A-Posteriori solution achieved by simulated annealing. In the nature of things, more than one convergent subsequence may be expected to exist
that yields this same maximal estimate, with the same asymptotic annealing temperature $T^{\infty}_{a}$. 
We ignore all subsequences except these maximal ones.

In References \cite{Morgan2005} and \cite{Morgan2012} expectation values for $T$ and  $\{\epsilon(k)\}$ over the the posterior probability Eqn. (\ref{eqn3}) were shown to give good estimates for physical surface temperatures and emissivities. We claim that the mean of a large number of subsequences that converge to the limiting MAP value will tend to the expectation values for $T$ and  $\{\epsilon(k)\}$ with respect to Eqn. (\ref{eqn3}).

The MAXENT estimator is constructed from the posterior probability of noise power in a spectral bin. For the sake of simplicity we assume identical noise power in each bin.\footnote{This assumption is inessential and may be relaxed.} 
A fully annealed MAP estimate may be thought of as an individual Bernoulli trial drawn from the likelihood function for $\{T_{n},\epsilon_{i}(k)\}$. By construction, all such trials are independent and identically distributed with bounded expectation values.\footnote{Moments over 
Eqn. (\ref{eqn3}) are 
bounded despite bad behavior of the Jeffreys prior at T=0, because of the rapid decay of the exponentials away from the MAP solution, as 
$L'H\hat{o}pital's$ rule demonstrates.} 

Let
\begin{equation}
\overline{T }=\frac{1}{N} \sum_{i=1}^{N} T_{i}
\end{equation} 
and
\begin{equation}
\overline{\epsilon(k)} =\frac{1}{N}  \sum_{i=1}^{N} \epsilon_{i}(k)
\end{equation} 
be the means of MAP surface temperature and spectral emissivity taken over 
 over $N$ convergent subsequences.
Suppose the covariance matrix $\mathbf{\Sigma}$ 
of the trials to be nonsingular. We invoke the multivariate Central Limit Theorem to conclude the mean values converge weakly to the multivariate Gaussian distribution:
\begin{equation}
\sqrt{N}  \left ( \begin{array}{c}
\overline{T }- \langle T \rangle \\  
\overline{\epsilon(1)}-\langle \epsilon(1) \rangle \\ \label{eq:CLT}
\vdots \\
\overline{\epsilon(m)}-\langle \epsilon(m) \rangle
\end{array} \right ) \leadsto \mathbf{N}_{m}(0,\mathbf{\Sigma}).
\end{equation} 
Reliance on the mixing hypothesis in the form given by Eq. (\ref{eq:mixed}), however, brings with it the concern that the relevant covariance matrix might be singular. 
In that event, the strong law of large numbers \cite{VanDerWaart1998,ArtsteinVitale1975} ensures
\begin{equation}
\overline{T } \stackrel{a.s.}{\rightarrow}  \langle T \rangle
\end{equation} 
and
\begin{equation}
\overline{\epsilon(k)} \stackrel{a.s.}{\rightarrow} \langle \epsilon(k) \rangle,
\end{equation} 
but without giving estimated variances of the mean values, such as come with Eq. (\ref{eq:CLT}).

To the extent that the estimator used in the simulated annealing search is zero-mean error, we conclude the estimates
yield accurate values for the physical values of $T$ and $\{\epsilon(k)\}$. 
As the spectral weights $\mathbf{x}$, lying as they do between zero and unity, possess bounded moments, this conclusion applies to the limiting mean values of $\{T_{n},\{x^{n}_{m}\}\}$ as well.
%


\subsection{Arbitrarily fine spectral resolution}

It is worth considering briefly the limiting case of infinite spectral resolution for $\epsilon(k)$.
In this section only, $k$ refers to wavenumber. We shall admit as endmembers any continuous function $\epsilon_{i}(k)$ on a compact interval $[k_{1},k_{2}] \in \mathbf{R}$, with  
\begin{equation}
0 \le \epsilon_{i}(k) \le 1, \label{eq:bndemiss}
\end{equation}
and bounded first derivative
\begin{equation}
\left \|\frac{d \epsilon_{i}(k)}{dk}  \right \|_{sup}  < W \label{eq:bndderiv}
\end{equation}
on that interval.

Consider 
\begin{equation}
f_{m_{n}}(k)=\epsilon(k)-\sum_{j=0} ^{m_{n}}\lambda_{j} \epsilon_{j}(k), \label{eq:squeeze}
\end{equation}
where, as before, as $m_{n}$ increases without bound, an admixture of arbitrary $ \epsilon_{j}(k)$ is included in the spectral mixture.
A standard argument shows that Eqs. (\ref{eq:bndemiss}) and (\ref{eq:bndderiv}) imply the spectral emissivities and Eq. (\ref{eq:squeeze}) are all members of an equicontinuous set. By the Arzel\`{a}-Ascoli lemma, as $m_{n} \rightarrow \infty$ there is a subsequence of trial emissivity spectra
\begin{equation}
\epsilon^{t}(k)=\sum_{j=0} ^{m_{n}}\lambda_{j} \epsilon_{j}(k)
\end{equation} 
for which an $M$ exists
such that for $m_{n} > M$ and for any positive $\delta$
\begin{equation}
| f_{m_{n}}(k)|  < \delta
\end{equation}
uniformly on $[k_{1},k_{2}]$. 
Thus, any physically admissible spectral emissivity $\epsilon(k)$ may be approximated arbitrarily well by a suitable spectral mixture of an unlimited number of
end members. 

We may regard the spectral mixture $x$ as an upper function on $[0,1] $ and take for a norm the Lebesgue
measure in the limit $m_{n} \rightarrow \infty$, with associated metric 
\begin{equation}
d(x,y)=\int d \mu (x-y).
\end{equation}
A mixture $x$ then becomes a vector in a Banach space.
As $n \rightarrow \infty$ the sequence of limiting fixed points under the 
self-mapping $\Phi$ 
has a convergent subsequence
$x_{n}$ which, again, satisfies the Cauchy condition
\begin{equation}
d(x_{n},x_{n+1}) < \epsilon
\end{equation}
The argument from the Cauchy property of the convergent subsequence of $x_{n}$ to the conclusion that the subsequence converges to a MAP equilibrium point
likewise follows much as before. 
As $n \rightarrow \infty$
\begin{equation}
0 < p(x_{n+1})-p(x_{n}) < \epsilon,
\end{equation}
so that the convergent subsequence of annealing trials, again, tends to an equilibrium point.
\section{DISCUSSION}

In Section \ref{sect:convergence} we proved convergence of simulated annealing searches for candidate MAP TES solutions. In Section \ref{Uniqueness} we argued that the average of a large number of these candidate MAP solutions converges almost surely to a unique estimate of surface temperature and spectral emissivity that, given a forward model leading to an unbiassed estimator for $T$ and $\{ \epsilon_{k} \}$, closely approximates the true values of these quantities. 

The motivation for seeking the Nash equilibrium analogy came from the realization that spectral mixing theory \cite{JohnsonEtAl1985,AdamsEtAl1986,AdamsEtAlCorrigenda} amounts to the use of mappings into a simplex, and that iterative choice of weights in the TES problem according to an annealing schedule amounts to a self-mapping into a convex polytope. The analogy with spectral mixing theory, however, is incomplete: Spectral unmixing, in either its reflective or thermal variants, generally appends and extra end member called "virtual dark " or "virtual cold", used to accommodate the effect of contamination from noise and sensor artifacts. One may see the value of an extra end member by recalling that the highest order components in a principle components decomposition of a multivariate dataset tends to be dominated by noise and artifacts that do not correlate with the physical content of lower-order components. 
The role of a virtual garbage end member will be discussed in Part Two.

A special case of great interest is the situation in which one seeks evidence that spectral radiance sensed at the top of the atmosphere (TOA) contains evidence for the presence of a specific spectral component. This problem may be addressed by use of spectral mixtures comprised of the desired spectral component together with generic continuous end members whose number is allowed to grow without bound. The analysis of Section \ref{sect:arbitrary}, however, cannot simply be modified by appending an endmember corresponding to the desired emissivity spectrum, as any completely random choice of $\epsilon(k)$ may well be correlated with the stipulated spectrum and so introduce a spurious admixture of that endmember into the analysis. On the other hand, in a physical mixture of spectral endmembers such correlations might in fact occur naturally. Without going into details, we offer some thoughts on this problem.

One way to proceed is to randomly select $\epsilon_{i}(k)$ for the background in such a way that the otherwise arbitrary background emissivity spectra all lie in the null space of the stipulated endmember. The choice of background $\epsilon(k)$ is thus made in much the same way as an empirical orthogonal basis set is selected.

Depending on the intended application, another way to handle this problem might be to find some way of marginalizing on the set of arbitrary background $\epsilon(k)$ in the calculation of the posterior probability Eq. (\ref{eq:posterior}). If performed by the same sort of stochastic sampling typical of simulated annealing, this calculation would resemble a numerical approximation to the Feynman-Kac formula.\cite{Kac}; in essence, a Monte Carlo path integral calculation.



\section{CONCLUDING REMARKS}

The practical utility of the mathematical development in this paper may be questioned. We address briefly two possible concerns.

While convergence of the algorithm has been proved to our satisfaction, we have no equally satisfactory estimates of the rate of convergence, with the consequence that the choice of annealing schedule remains a matter of trial-and-error. In response to this concern, the availability of massively parallel computation made possible by the ready availability of cheap GPU arrays means that massive processing requirements need not preclude the use of a resource-hungry algorithm if that algorithm can provide performance not attainable by other approaches. Part Two of this study will address these issues.

Another legitimate concern is that the spectral emissivity of natural ground covers in the wild will seldom be known to the level of accuracy found in Ref. \cite{Salisbury1992}. While true in general, this concern has not dissuaded other researchers from relying upon spectral unmixing.

The simulated annealing approach to TES by spectral unmixing does, however, offer something that other TES algorithms do not: By construction, it gives (in the limit) the unique best estimate in a Maximum \emph{A-Posteriori} sense, for the remote determination of surface temperature and spectral emissivity of a patch of ground that is known to be comprised of a spectral mixture of a stipulated set of spectral end members.


\nocite{*}
\bibliographystyle{IEEE}

John Morgan is a Senior Engineering Specialist in the Sensing and Exploitation Department at The Aerospace Corporation, where his duties include studies of spacecraft sensor system performance and remote sensing phenomenology.  He has a BS from Caltech, and MS and Ph.D. degrees from Rice University, in physics.

\end{document}